\documentclass[prd,aps,twocolumn,nofootinbib,preprintnumbers,superscriptaddress,preprintnumbers,balancelastpage,longbibliography]{revtex4-2}

\usepackage{graphicx}
\usepackage{bm}
\usepackage{hyperref}
\usepackage{slashed}
\usepackage{comment}
\usepackage{aas_macros}
\usepackage{float}
\usepackage{slashed}
\usepackage{lipsum}
\usepackage{subfigure}
\usepackage{multirow}
\usepackage{amsmath}
\usepackage{array} 
\usepackage{varwidth} 
\usepackage{tabularx}
\usepackage{braket}
\usepackage[dvipsnames]{xcolor}

\usepackage[normalem]{ulem}

\newcommand{\be}{\begin{equation}}
\newcommand{\ee}{\end{equation}}
\newcommand{\bea}{\begin{eqnarray}}
\newcommand{\eea}{\end{eqnarray}}

\hypersetup{
     colorlinks   = true,
     citecolor    = teal,
     urlcolor     = teal,
     linkcolor    = teal
}

\usepackage{listings}
\usepackage{color,xcolor}

\begin{document}

\title{Constraints and Projections for Millicharged Dark Matter in the Sun with Water Cherenkov Neutrino Detectors}

\author{Thong T.Q. Nguyen}
\thanks{{\scriptsize Email}: \href{mailto:thong.nguyen@fysik.su.se}{thong.nguyen@fysik.su.se}; \href{https://orcid.org/0000-0002-8460-0219}{0000-0002-8460-0219}}
\affiliation{Stockholm University and The Oskar Klein Centre for Cosmoparticle Physics, Alba Nova, 10691 Stockholm, Sweden}

\begin{abstract}
Millicharged particles are well-motivated dark matter candidates that have been extensively investigated in terrestrial experiments. Recent studies proposed using the IceCube Neutrino Observatory to search for high-energy neutrinos produced by the capture and annihilation of millicharged dark matter in the Sun, deriving new constraints in the strong interaction regime where the millicharge is $q_\chi \sim 10^{-3}$--$10^{-2}$, which extend to small fractional abundances where all existing constraints lose sensitivity. In this work, I point out that the lower energy threshold of water Cherenkov detectors makes Super-Kamiokande and the future Hyper-Kamiokande sensitive to neutrinos from the annihilation of lighter millicharged dark matter, complementing the high-mass reach of IceCube. I find that Super-Kamiokande can constrain previously unexplored parameter space at $m_{\chi}=5$--28~GeV for dark matter fraction of $f_{\chi}=10^{-4.5}$, while Hyper-Kamiokande can improve these constraints and will be sensitive to fractional abundances as small as $f_{\chi}\simeq 5\times 10^{-6}$, nearly an order of magnitude below current IceCube limits.
\end{abstract}

\maketitle

\section{Introduction}
\label{sect:intro}

Although there is overwhelming evidence for the existence of dark matter (DM)~\cite{Bertone:2016nfn}, identifying how DM interacts with the Standard Model (SM) non-gravitationally remains one of the central missions of modern physics~\cite{Bertone:2004pz, Bertone:2018krk, Cirelli:2024ssz}. Among the proposed Beyond the Standard Model (BSM) candidates, millicharged DM stands out as a particularly well-motivated possibility. The simplest realization extends the SM by an additional $U(1)^{\prime}$ gauge symmetry that kinetically mixes with the SM hypercharge. In the limit where this dark $U(1)^{\prime}$ remains unbroken, or equivalently where the dark photon $A^{\prime}$ is sufficiently light that its mass plays no role at the relevant momentum scales~\cite{Berlin:2022hmt}, the dark-sector field $\chi$ acquires a small effective electric charge under the SM electromagnetic gauge group, $q_{\chi}$, allowing it to interact with SM particles through long-range Coulomb-like exchanges of the ultralight dark photon~\cite{Goldberg:1986nk, Feng:2009mn, DeRujula:1989fe}.

A rich experimental and theoretical program has been developed to search for this charged DM candidate~\cite{Haas:2014dda, Izaguirre:2015eya, Magill:2018tbb, Kelly:2018brz, ArgoNeuT:2019ckq, Harnik:2019zee, Oscura:2023qch, Plestid:2020kdm, Bloch:2020uzh, Foroughi-Abari:2020qar, Harnik:2020ugb, Marocco:2020dqu, Berlin:2023gvx, Chen:2022abz, Budker:2021quh, ArguellesDelgado:2021lek, Afek:2020lek, Berlin:2025btf, Berlin:2025hjs, Berlin:2024ewa, Berlin:2024dwg, Berlin:2021kcm, Berlin:2020pey, Prabhu:2022dtm, Du:2024afd, Essig:2024dpa}. However, when millicharged particles make up only a small fraction of the total DM relic abundance, constraints from the cosmic microwave background (CMB)~\cite{Kovetz:2018zan, dePutter:2018xte, Munoz:2018pzp, Creque-Sarbinowski:2019mcm} and big bang nucleosynthesis (BBN)~\cite{Stebbins:2019xjr}, as well as those from balloon-borne~\cite{Rich:1987st} and space-based~\cite{Erickcek:2007jv} experiments, all weaken substantially. This leaves an open window between the direct-detection ceiling~\cite{Emken:2019tni} and collider bounds~\cite{CMS:2012xi, Davidson:2000hf, CMS:2013czn, CMS:2024eyx}, requiring different strategies to probe this strongly-coupled parameter space.

Recently, Berlin and Hooper~\cite{Berlin:2024lwe} have proposed a novel approach: the Sun can accumulate a large population of millicharged particles through its gravitational potential. They investigate how the captured millicharged DM forms bound states with solar nuclei and annihilates into SM final states that produce a high-energy neutrino flux at Earth. Using the leading sensitivity to high-energy solar neutrinos from IceCube~\cite{IceCube:2021xzo}, Ref.~\cite{Berlin:2024lwe} sets strong constraints reaching down to $f_{\chi}\simeq 5\times 10^{-5}$ for DM masses in the $30\text{--}100$~GeV range, excluding the unexplored millicharges around $q_{\chi}\sim 10^{-3}$.

Building on this framework, in this paper I study the high-energy muon neutrino flux produced by millicharged DM annihilation inside the Sun, using a 10-year observation timescale of atmospheric muon neutrinos at water Cherenkov detectors, namely Super-Kamiokande (Super-K) and the upcoming Hyper-Kamiokande (Hyper-K). The lower energy threshold of these detectors allows the analysis to extend down to $m_{\chi}=2$~GeV, due to the evaporation suppression in the strong DM-baryon scattering regime inside the Sun. For the smallest DM fraction probed by IceCube in Ref.~\cite{Berlin:2024lwe}, I show that Super-K can cover previously unexplored regions of the $q_{\chi}$--$m_{\chi}$ parameter space between $5\text{--}28$~GeV DM masses, while the projected reach of Hyper-K bridges the gap between Super-K and IceCube. Hyper-K is further projected to probe DM fractions an order of magnitude below the current IceCube limit. Several natural follow-ups to this direction are discussed at the end of this work.

This paper is organized as follows. Section~\ref{sect:model} reviews the millicharged DM model. Section~\ref{sect:Solar_mCP} discusses the physics of millicharged DM accumulation, distribution, bound-state formation with solar nuclei at large millicharge, and annihilation inside the Sun. Section~\ref{sect:flux} describes solar neutrino observations at water Cherenkov detectors, including both the atmospheric background and the signal from DM annihilation. In Section~\ref{sect:constraint}, I present the Super-K constraints and Hyper-K projections, and compare them with previous studies, in particular the IceCube results of Ref.~\cite{Berlin:2024lwe}. I conclude in Section~\ref{sect:conclusion} and outline several future directions for the millicharged DM solar capture scenario.

\section{Millicharged Dark Matter Model}
\label{sect:model}

Millicharged DM typically arises in extensions of the Standard Model (SM) with an additional $U(1)^{\prime}$ gauge symmetry, and interacts with SM particles through an ultralight dark photon $A^{\prime}$ that kinetically mixes with the SM photon~\cite{Fabbrichesi:2020wbt, Fayet:1980rr, Holdom:1985ag}. The Lagrangian reads
\begin{equation}
\begin{split}
    \mathcal{L}\supset &-\frac{1}{4}F_{\mu\nu}^{\prime}F^{\prime \mu\nu}-\frac{\epsilon}{2}F_{\mu\nu}F^{\prime \mu\nu}\\
    &+\frac{1}{2}m_{A^{\prime}}^{2}A_{\mu}^{\prime}A^{\prime \mu} + \bar{\chi}(i\slashed{D}-m_{\chi})\chi,
    \label{eq:lagrangian}
\end{split}
\end{equation}
where the covariant derivative acting on the DM field $\chi$ is
\be
    D_{\mu}=\partial_{\mu}-i e^{\prime} A^{\prime}_{\mu},
    \label{eq:Dmu}
\ee
with $e^{\prime}$ the coupling between the dark photon and the millicharged particle. The effective millicharge of the DM particle is given by 
\be
q_{\chi}=\epsilon\, e^{\prime} / e\,,
\label{eq:qX}
\ee
where $\epsilon$ is the kinetic mixing parameter. A one-loop diagram involving heavy particles charged under both the SM hypercharge $U(1)$ and the dark $U(1)^{\prime}$ generates this mixing with size
\begin{equation}
    \epsilon \sim \frac{e^{\prime}e}{16\pi^{2}}\ln\left(\frac{M^{\prime}}{M}\right)\simeq 6\times 10^{-4}\, \left(\frac{e^{\prime}}{e}\right)\ln\left(\frac{M^{\prime}}{M}\right),
    \label{eq:kineticmixing}
\end{equation}
where $M$ and $M^{\prime}$ are the masses of the heavy particles running in the loop~\cite{Cohen:2010kn, Holdom:1985ag, Baumgart:2009tn}. For $e^{\prime}\sim e$, this naturally yields $q_{\chi}\sim 10^{-3}$. In this work, however, I treat $q_{\chi}$ and the dark fine-structure constant $\alpha_{D}=(e^{\prime})^{2}/4\pi$ as independent parameters, focusing on the regime $\alpha_{D}\ll \alpha$ where annihilation into dark photons is suppressed and SM final states dominate (see Section~\ref{ssect:Ann}).

If millicharged particles are a dark matter candidate, cosmological observations require them to be a subdominant component rather than to make up the full relic abundance. Measurements of the cosmic microwave background and the primordial light-element abundances constrain the charged DM fraction to $f_{\chi}\lesssim 10^{-2}$~\cite{Dubovsky:2003yn, dePutter:2018xte, Kovetz:2018zes, Buen-Abad:2021mvc, Stebbins:2019xjr}, while perturbative unitarity of thermal freeze-out sets a lower limit
\be
    f_{\chi}\gtrsim 10^{-10}\times (m_{\chi}/\text{GeV})^{2}~,
    \label{eq:fX}
\ee
for a thermal relic millicharged subcomponent~\cite{Griest:1989wd}. Within this window, underground direct detection experiments rule out charges as small as $q_{\chi}\sim 10^{-10}/f_{\chi}^{1/2}$~\cite{Emken:2019tni}, but lose sensitivity at $q_{\chi}\gtrsim 10^{-4}\times(m_{\chi}/\text{GeV})^{1/2}$, where millicharged particles thermalize in the terrestrial overburden well below detector thresholds~\cite{Emken:2019tni}.

This blind spot at moderate to large couplings has motivated a rich program of complementary searches. These include collider searches by the CMS collaboration~\cite{CMS:2012xi, Davidson:2000hf, CMS:2013czn, CMS:2024eyx}, the X-ray Quantum Calorimeter (XQC)~\cite{Erickcek:2007jv, Mahdawi:2018euy}, the balloon-based Rich-Rocchia-Spiro (RRS) experiment~\cite{Rich:1987st}, searches for molecular cloud ionization by DM scattering~\cite{Prabhu:2022dtm, Blanco:2023bgz}, and constraints from the James Webb Space Telescope (JWST)~\cite{Du:2024afd}. In this work, I study the scenario where millicharged DM accumulates inside the Sun and annihilates to produce neutrinos, which was proposed by Ref.~\cite{Berlin:2024lwe}, and can be tested by water Cherenkov detectors.

\section{Millicharged Dark Matter in the Sun}
\label{sect:Solar_mCP}

With its large radius, $R_{\odot}\simeq 6.957\times 10^{5}$~km, and age, $t_{\odot}\simeq 4.6$~Gyr, the Sun acts as a natural DM detector, providing a gigantic target for DM scattering. First proposed in 1985 by Refs.~\cite{Press:1985ug, Faulkner:1985rm, Silk:1985ax, Srednicki:1986vj, Gaisser:1986ha}, the basic idea is that incoming DM particles from the Galactic halo can fall into the solar gravitational potential and scatter with nucleons and electrons inside the solar medium, losing kinetic energy and becoming gravitationally bound. This accumulation enhances the DM density inside the Sun, boosting the annihilation rate and opening a variety of indirect-detection strategies based on solar neutrino and $\gamma$-ray observations~ \cite{IceCube:2025fcu, Hooper:2025ohk, Berlin:2024lwe, ANTARES:2016obx, Widmark:2017yvd, Catena:2016ckl, IceCube:2016yoy, Acevedo:2020gro, Ng:1986qt, Krishna:2025ncv, Maity:2023rez, IceCube:2021xzo, Bell:2021esh, Bell:2011sn, Bell:2012dk, Kopp:2009et, Super-Kamiokande:2015xms, Leane:2017vag, HAWC:2018szf, HAWC:2022khj, Nisa:2019mpb, HAWC:2018szf, Bose:2021cou, Bell:2021pyy, Serini:2022aed, FermiLAT:2011ozd, Cuoco:2019mlb}.

\subsection{Millicharged dark matter accumulation, distribution, and evaporation}
\label{ssect:capt_nX_eva}

In the strongly-coupled regime, the cross section between millicharged DM and baryonic matter exceeds the Sun's transition cross section~\cite{Leane:2023woh}, previously referred to as the ``saturation" cross section in the literature~\cite{Bramante:2017xlb, Leane:2021ihh, Nguyen:2022zwb, Ilie:2024sos, Bell:2021fye}, with a value of $\sigma_{\rm tr}\simeq 10^{-35}$~cm$^{2}$. The Sun has a large escape velocity, ranging from $\sim 1400$~km/s near the core to $\sim 600$~km/s at the surface~\cite{Nguyen:2025ygc}, well above the average DM velocity in the solar neighborhood, $v_{\rm DM}\simeq 270$~km/s. In this regime, where the DM-nucleon cross section exceeds $\sigma_{\rm tr}$, the Sun becomes optically thick and every millicharged DM particle passing through it is captured~\cite{Leane:2023woh}. The effective capture area is~\cite{Berlin:2024lwe}
\begin{equation}
    A_{\odot}=\pi R_{\odot}^{2}\left( 1 + \frac{v_{\rm esc}^{2}}{v_{\rm DM}^{2}} \right),
\end{equation}
where $v_{\rm esc}$ is taken at the solar surface. The corresponding capture rate is
\begin{equation}
    \Gamma_{\rm cap}\simeq f_{\chi}\, \frac{\rho_{\chi}}{m_{\chi}}\, v_{\rm DM}\, A_{\odot},
\end{equation}
where $\rho_{\chi}=0.4$~GeV/cm$^{3}$ is the local DM density~\cite{Salucci:2010qr}, and $f_{\chi}$ is the DM fraction.

Once trapped inside the Sun, millicharged DM particles distribute themselves under the influence of the solar gravitational potential. To find their distribution, Ref.~\cite{Berlin:2024lwe} solves the equation arising from the hydrostatic equilibrium condition
\begin{equation}
    \frac{1}{n_{\chi}(r)}\frac{{\rm d}n_{\chi}(r)}{{\rm d}r} + \frac{1}{T_{\odot}(r)}\frac{{\rm d}T_{\odot}(r)}{{\rm d}r}+\frac{m_{\chi}\,g_{\odot}(r)}{T_{\odot}(r)}\simeq 0,
    \label{eq:PDE_hydro}
\end{equation}
where $T_{\odot}(r)$ is the solar temperature profile and the gravitational acceleration acting on trapped DM particles is
\begin{equation}
    g_{\odot}(r) = \int_{0}^{r}\frac{G_{N}\,M_{\odot}(r^{\prime})}{r^{\prime 2}}{\rm d}r^{\prime},
\end{equation}
with $M_{\odot}(r)$ the solar mass enclosed within radius $r$. Following Ref.~\cite{Berlin:2024lwe}, I adopt the Standard Solar Model for the solar mass density and temperature profiles~\cite{Magg:2022rxb}. However, instead of solving Eq.~\ref{eq:PDE_hydro}, I exploit the fact that in the strongly-interacting regime, the dense solar interior is optically thick to DM scattering, so the trapped DM particles thermalize locally with the surrounding plasma. While the optically thick condition breaks down in the tenuous outer layers, this has a negligible effect on the captured population: for $m_{\chi}\geq 2$~GeV, the thermalized DM is strongly concentrated within the inner few percent of the solar radius~\cite{Berlin:2024lwe}, where the optically thick approximation is well satisfied. Setting the DM thermal temperature $T_{\chi}(r)=T_{\odot}(r)$, the DM radial distribution can then be approximated as\cite{Banks:2024eag, Nauenberg:1986em, Gould:1989hm}

\begin{align}
    n_{\chi}^{\rm LTE}(r) =\ & N_{0}^{\rm LTE}\left( \frac{T_{\odot}(r)}{T_{\odot}(0)} \right)^{3/2}  \\
    & \times \exp\left( -\int_{0}^{r}\frac{\alpha(m_{\chi})\frac{{\rm d}T_{\odot}(r^{\prime})}{{\rm d}r^{\prime}} + m_{\chi}\,g_{\odot}(r^{\prime})}{T_{\odot}(r^{\prime})}\,{\rm d}r^{\prime} \right),\nonumber
\end{align}
where the normalization factor $N_{0}^{\rm LTE}$ is fixed by the condition $\int_{0}^{R_{\odot}}n_{\chi}^{\rm LTE}(r)\,4\pi r^{2}\,{\rm d}r = 1$. I adopt the thermal diffusion coefficient $\alpha(m_{\chi})$ from Refs.~\cite{Gould:1989hm, Leane:2022hkk, Gould:1989tu}. The normalized annihilation rate, which will enter the DM annihilation calculation below, is given by
\begin{equation}
    \frac{\langle n_{\chi}^{2}\rangle}{\langle n_{\chi}\rangle^{2}}=\frac{\int_{0}^{R_{\odot}}4\pi r^{2}\,n_{\chi}^{2}(r)\,{\rm d}r}{\left(\int_{0}^{R_{\odot}}4\pi r^{2}\,n_{\chi}(r)\,{\rm d}r\right)^{2}},
\end{equation}
where I evaluate $n_{\chi}(r)$ using the LTE approximation. This result agrees with the calculation in Ref.~\cite{Berlin:2024lwe}.

After being distributed inside the Sun, captured DM particles can also be up-scattered by hot moving electrons and positrons through the same interaction that drives the capture process. For light DM, if the kinetic energy injected by an electron or positron is large enough to boost the DM velocity above the solar escape velocity, the DM particle escapes the Sun. This process, known as DM evaporation, suppresses the number of captured DM particles~\cite{Gould:1987ju}. For solar DM capture, evaporation becomes relevant for DM masses below $\sim4$~GeV~\cite{Garani:2021feo}, and previous studies of indirect detection signals from solar DM capture therefore typically impose a lower mass cutoff around 4--5~GeV~\cite{IceCube:2025fcu, Hooper:2025ohk, Berlin:2024lwe, ANTARES:2016obx, Widmark:2017yvd, Catena:2016ckl, IceCube:2016yoy, Acevedo:2020gro, Ng:1986qt, Krishna:2025ncv, Maity:2023rez, IceCube:2021xzo, Bell:2021esh, Bell:2011sn, Bell:2012dk, Kopp:2009et, Super-Kamiokande:2015xms, Leane:2017vag, HAWC:2018szf, HAWC:2022khj, Nisa:2019mpb, HAWC:2018szf, Bose:2021cou, Bell:2021pyy, Serini:2022aed, FermiLAT:2011ozd, Cuoco:2019mlb}.

However, in the parameter space considered in this work, where the cross section between millicharged DM and proton/electron exceeds the transition cross section threshold, the DM thermalizes locally and reaches local thermal equilibrium (LTE) with the surrounding plasma. In this strongly-coupled regime, evaporation is suppressed: even when an upscattering event imparts enough energy for the DM to escape, the particle undergoes additional scatterings on its way out of the Sun, losing kinetic energy and remaining trapped~\cite{Gould:1989tu}. Refs.~\cite{Garani:2017jcj, Busoni:2017mhe, Nguyen:2026nhe} have explicitly computed this evaporation suppression and shown that the DM evaporation mass can be pushed down to $\sim 2$~GeV. Motivated by this, I extend the millicharged DM capture calculation down to $m_\chi = 2$~GeV, whereas the previous study of Ref.~\cite{Berlin:2024lwe} stopped at the traditional 5~GeV cutoff~\footnote{A recent study, Ref.~\cite{Acevedo:2023owd}, points out that long-range attractive forces between DM and SM particles (corresponding to an ultralight mediator) can further push the evaporation barrier to even smaller DM masses. Adapting this treatment to the millicharged case is nontrivial, since the ultralight dark photon mediates both attractive and repulsive long-range Coulomb forces between $\chi^{\pm}$ and SM matter. I do not include this effect in the present work, and conservatively adopt the standard $\sim 2$~GeV evaporation mass~\cite{Garani:2017jcj, Busoni:2017mhe} as the lower limit of my analysis.}.

\subsection{Millicharged particle-nucleus bound state}
\label{ssect:boundstate}

Millicharged DM consists of both positively and negatively charged states, $\chi^{+}$ and $\chi^{-}$, which are captured by the Sun through the same scattering process. One caveat of this study is that I assume the Sun captures equal populations of $\chi^{+}$ and $\chi^{-}$, so that the trapped DM forms a charge-symmetric ensemble.

The presence of both species opens a new channel that affects the annihilation dynamics: the negatively charged $\chi^{-}$ can bind to atomic nuclei inside the Sun and form Coulomb bound states $(\chi^{-}N)$ with characteristic binding energy
\begin{equation}
    E_{\chi N}\simeq \frac{(q_{\chi}Z\alpha)^{2}\mu_{\chi N}}{2}\,,
\end{equation}
where $Z$ is the atomic number of the nucleus, $\alpha$ is the fine-structure constant, and $\mu_{\chi N}=m_{\chi}m_{N}/(m_{\chi}+m_{N})$ is the reduced mass. For sufficiently large $q_{\chi}$ and $Z$, $E_{\chi N}$ can substantially exceed the solar core temperature $T_{\odot}\simeq 1$~keV, and bound-state formation becomes energetically favored.

Once bound, the $(\chi^{-}N)$ system carries a net positive charge $Q_{\chi N}=Z-q_{\chi}\simeq Z$. Annihilation between $\chi^{+}$ and the bound $\chi^{-}$ is then suppressed by the repulsive Coulomb barrier between the incoming $\chi^{+}$ and the net positive charge $Q_{\chi N}$ of the bound system. When the binding energy $E_{\chi N}\gtrsim T_{\odot}$, the colliding particles do not have enough thermal kinetic energy to overcome this repulsion, and the annihilation rate is exponentially suppressed~\cite{Pospelov:2020ktu, Berlin:2024lwe}.

Assuming that the formation and disruption of bound states are rapid enough to maintain chemical equilibrium between the bound $(\chi^{-}N)$ and free $\chi^{-}$ populations, the Maxwell-Boltzmann statistics give
\begin{equation}
    \frac{n_{\chi^{-}}\times n_{N}}{n_{(\chi^{-}N)}}=\left( \frac{\mu_{\chi N}T_{\odot}}{2\pi} \right)^{3/2}e^{-E_{\chi N}/T_{\odot}}\,.
\end{equation}
The effective annihilation cross section of millicharged DM to any Standard Model final state is then rescaled as
\begin{equation}
    \langle \sigma v\rangle_{\chi^{+}\chi^{-}\to {\rm SM}}\to R_{\rm BS}(m_{\chi}, q_{\chi})\times\langle \sigma v\rangle_{\chi^{+}\chi^{-}\to {\rm SM}},
\end{equation}
where the bound-state suppression factor is
\begin{equation}
    \begin{split}
    R_{\rm BS}&=\frac{F_{N}+1}{F_{N}+e^{E_{\chi N}/T_{\odot}}},\\ F_{N}&= \left(\frac{\mu_{\chi N}T_{\odot}}{2\pi}\right)^{3/2}\frac{1}{n_{N}}.
    \end{split}
\end{equation}
The annihilation rate is thus exponentially suppressed when $E_{\chi N}\gtrsim T_{\odot}\ln F_{N}$.

To remain robust and conservative, I follow Ref.~\cite{Berlin:2024lwe} and consider bound-state formation only with thorium ($Z=90$), the highest-$Z$ element measured in the solar photosphere, with mass fraction \mbox{$\sim 2\times 10^{-10}$~\cite{Asplund:2009fu}}. Taking the corresponding number density \mbox{$n_{N}\simeq 7\times 10^{11}~\text{cm}^{-3}$} to be representative of the entire solar volume gives $\ln F_{N}\sim 50$ for $m_{\chi}\gtrsim 1$~GeV. Choosing thorium yields the most conservative (i.e., weakest) limits on $q_{\chi}$, since heavier nuclei produce larger binding energies and earlier onset of bound-state suppression. While lighter, more abundant elements such as oxygen would extend the excluded region to larger $q_{\chi}$, as shown in Ref.~\cite{Berlin:2024lwe}, restricting the analysis to thorium ensures that the resulting bounds are conservative.

\subsection{Millicharged dark matter annihilation}
\label{ssect:Ann}

Millicharged DM annihilates into Standard Model fermions through $s$-channel exchange of an ultralight dark photon kinetically mixed with the SM photon. In the non-relativistic limit, the thermally averaged annihilation cross section into a fermion pair with $N_{c}$ colors and electric charge $Q_{f}$ is~\cite{Berlin:2024lwe}
\begin{equation}
\begin{split}
    \langle \sigma v \rangle_{\chi^{+}\chi^{-}\to f\bar{f}}&\simeq \frac{\pi \alpha^{2} q_{\chi}^{2}}{m_{\chi}^{2}}\,N_{c}\,Q_{f}^{2}\\
    &\times \sqrt{1 - \frac{m_{f}^{2}}{m_{\chi}^{2}}}\,\left( 1 + \frac{m_{f}^{2}}{2m_{\chi}^{2}}\right)\,,
\end{split}
\end{equation}
where the last two factors include the phase-space and matrix-element corrections, which become important for $m_{\chi}$ near the bottom-quark mass.

For $m_{\chi}\gtrsim m_{W}$, millicharged DM can also annihilate directly into $W^{+}W^{-}$ pairs through the SM $\gamma WW$ coupling, as well as $ZWW$ if the dark photon kinetically mixes with both neutral gauge bosons~\cite{Nguyen:2024kwy}. The resulting $W$ decays contribute to a neutrino flux, which is observable by Super-K, Hyper-K, and IceCube. However, since the branching fraction into $W^{+}W^{-}$ is generally smaller than into charged fermion pairs~\cite{Nguyen:2024kwy}, and becomes negligible at the small DM fractions of primary interest in this work, I neglect this channel in what follows.

Since the interaction is mediated by an ultralight dark photon, millicharged DM can also annihilate directly into a pair of dark photons. The cross section for this process is~\cite{Berlin:2024lwe}
\begin{equation}
    \langle \sigma v \rangle_{\chi^{+}\chi^{-}\to A^{\prime}A^{\prime}}\simeq \frac{\pi \alpha_{D}^{2}}{m_{\chi}^{2}}\,,
\end{equation}
where $\alpha_{D}=e^{\prime 2}/4\pi$ is the dark fine-structure constant. This channel opens up new search avenues, such as gamma-ray observations through the dark photon trident decay $A^{\prime}\to 3\gamma$ outside the Sun~\cite{Linden:2024uph, Linden:2024fby, McDermott:2017qcg}. However, following Ref.~\cite{Berlin:2024lwe}, I focus on the limit of small $e^{\prime}$, where the dark photon channel is subdominant compared to annihilation into SM charged fermions.

\begin{figure*}[t!]
\centering
\includegraphics[width=1\columnwidth]{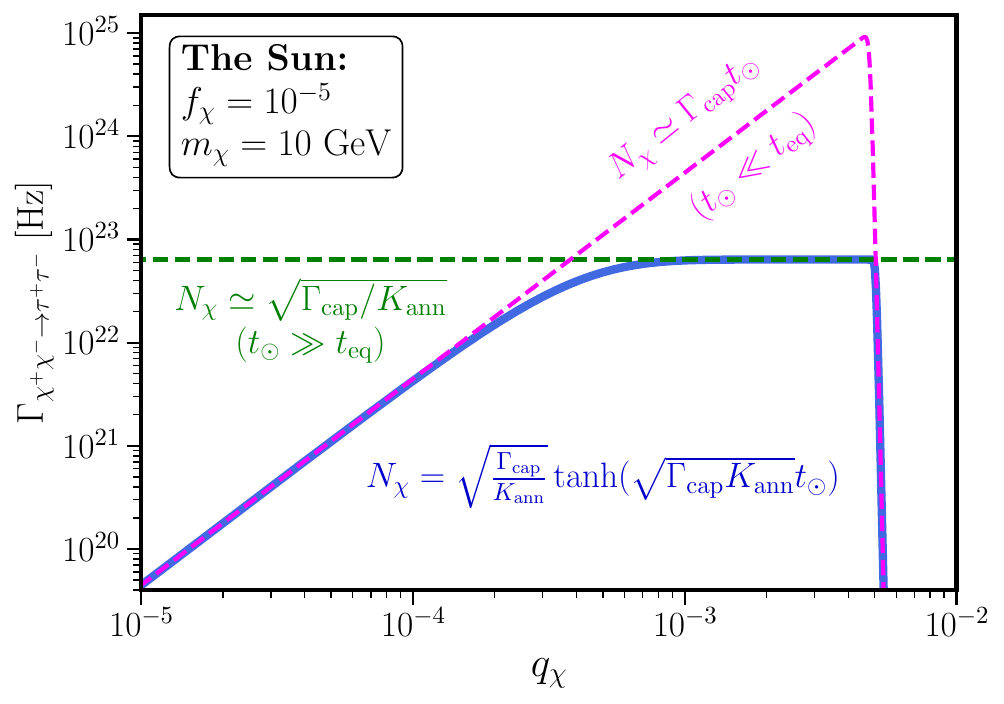}
\hfill
\includegraphics[width=1\columnwidth]{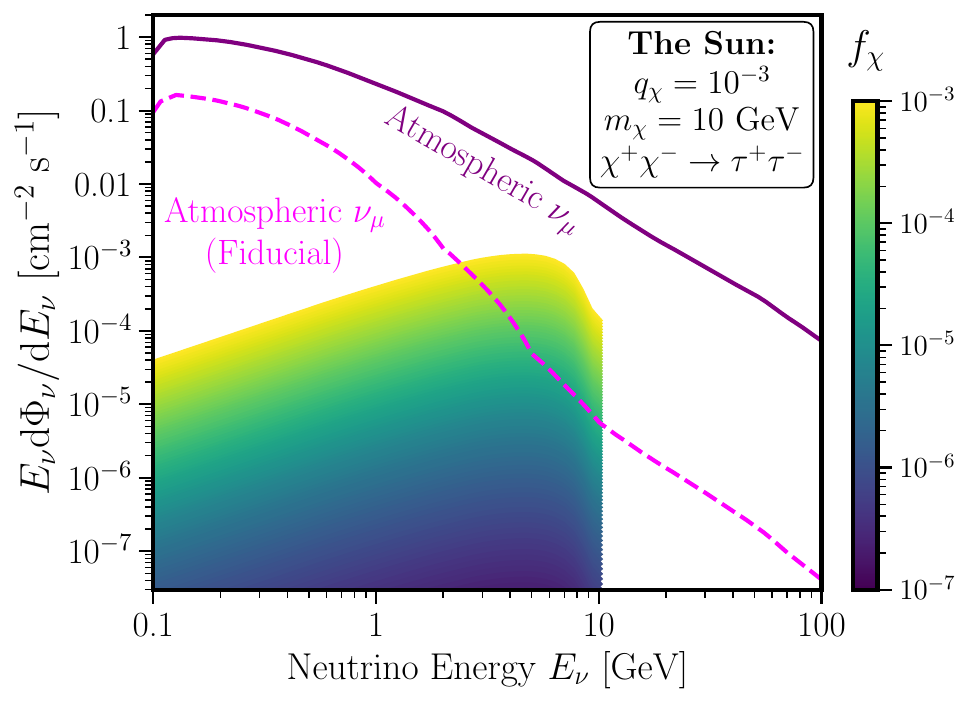}
\vspace{-0.5cm}
\caption{Millicharged DM with a benchmark mass $m_{\chi}=10$~GeV annihilating to $\tau^{+}\tau^{-}$ inside the Sun. {\bf Left:} DM annihilation rate as a function of the millicharge $q_{\chi}$ for a DM fraction of $f_\chi = 10^{-5}$, taking into account bound-state formation with Thorium nuclei in the Sun. The dashed lines show two approximations for the number of captured DM particles: the equilibrium limit (green) and the non-equilibrium limit (magenta). The solid blue line shows this work's full calculation that interpolates between both regimes. {\bf Right:} Muon neutrino fluxes from the atmospheric background (solid purple), and its fiducial component reduced by the angular resolution of Super-K/Hyper-K (dashed magenta). The neutrino fluxes from $\chi^{+}\chi^{-}\to\tau^{+}\tau^{-}$ annihilation for a benchmark millicharge $q_{\chi}=10^{-3}$ are shown for different DM fractions, indicated by the color bar.}
\label{fig:Obser}
\end{figure*}

The annihilation rate per pair of $\chi^{+}\chi^{-}$ inside the Sun into a specific SM final state is given by
\begin{equation}
\begin{split}
    K_{\rm ann}&=\frac{1}{2N_{\chi}^{2}}\int_{V_{\odot}}n_{\chi}^{2}\,\langle \sigma v \rangle_{\chi^{+}\chi^{-}\to {\rm SM}}\,{\rm d}V\\
    &\simeq \frac{\langle \sigma v \rangle_{\chi^{+}\chi^{-}\to {\rm SM}}}{2\, V_{\odot}}\,\frac{\langle n_{\chi}^{2}\rangle}{\langle n_{\chi}\rangle^{2}}\,,
\end{split}
\end{equation}
where $N_{\chi}$ is the total number of captured DM particles. For DM masses where evaporation is suppressed, the present-day number of captured millicharged DM particles is~\cite{Berlin:2024lwe}
\begin{equation}
    N_{\chi}\simeq \sqrt{\frac{\Gamma_{\rm cap}}{K_{\rm ann}^{\rm tot}}}\,\tanh\left( \sqrt{\Gamma_{\rm cap}\,K_{\rm ann}^{\rm tot}}\,\, t_{\odot} \right)\,,
\end{equation}
which has two limiting behaviors,
\begin{equation}
    N_{\chi}\simeq \Gamma_{\rm cap}\,t_{\odot}\quad(t_{\odot} \ll t_{\rm eq}:\text{ non-equilibrium})\,,
\end{equation}
\begin{equation}
    N_{\chi}\simeq \sqrt{\Gamma_{\rm cap}/K_{\rm ann}^{\rm tot}}\quad (t_{\odot}\gg t_{\rm eq}:\text{ equilibrium}),
\end{equation}
where $t_{\odot}\simeq 4.5$~Gyr is the age of the Sun, and
\begin{equation}
    t_{\rm eq} = 1/\sqrt{\Gamma_{\rm cap}\,K_{\rm ann}^{\rm tot}}
\end{equation}
is the timescale for DM to reach capture-annihilation equilibrium. The equilibration is governed by the total annihilation rate, which is calculated by summing over all kinematically accessible SM fermion channels,
\begin{equation}
    K_{\rm ann}^{\rm tot} = \sum_{f} K_{\rm ann}^{f\bar{f}}\,.
\end{equation}

For a specific annihilation channel relevant to indirect detection, in this case $\chi^{+}\chi^{-}\to \tau^{+}\tau^{-}$, the total annihilation rate is
\begin{equation}
    \Gamma_{\chi^{+}\chi^{-}\to\tau^{+}\tau^{-}}=\frac{1}{2}\,N_{\chi}^{2}\,K_{\rm ann}^{\tau^{+}\tau^{-}}\,.
\end{equation}
The left panel of Figure~\ref{fig:Obser} shows this rate as a function of the millicharge $q_{\chi}$, for a benchmark $m_{\chi}=10$~GeV (above the evaporation threshold) and a DM fraction of $f_{\chi}=10^{-5}$. The two limiting behaviors derived above are shown as dashed lines: the magenta line corresponds to the non-equilibrium limit, $N_{\chi}\simeq \Gamma_{\rm cap}t_{\odot}$, while the green line corresponds to the equilibrium limit, \mbox{$N_{\chi}\simeq \sqrt{\Gamma_{\rm cap}/K_{\rm ann}^{\rm tot}}$}. The full calculation (solid blue) smoothly interpolates between them. At small $q_{\chi}$, the annihilation cross section is small and the equilibration timescale exceeds the age of the Sun, so the DM remains in the non-equilibrium regime where the captured population grows linearly in time. As $q_{\chi}$ increases, the annihilation rate becomes large enough for capture-annihilation equilibrium to be reached, and $\Gamma_{\chi^{+}\chi^{-}\to\tau^{+}\tau^{-}}$ saturates. Beyond a critical value of $q_{\chi}$, however, bound-state suppression sets in and rapidly cuts off the annihilation rate. This effect determines the upper boundary of the parameter space accessible to indirect detection: at sufficiently large $q_{\chi}$, the trapped $\chi^{-}$ population is locked into $(\chi^{-}N)$ bound states, and the resulting SM neutrino flux drops below detector sensitivity.

\section{Solar Neutrino observation from Water Cherenkov detectors}
\label{sect:flux}

The Super-K detector, located under Mount Ikeno in the Gifu Prefecture of Japan, has played a central role in neutrino physics for the past several decades~\cite{Super-Kamiokande:2002weg}. With a fiducial mass of 22.5~kton of pure water, Super-K observes neutrino interactions through the Cherenkov radiation emitted by charged secondary particles, and has provided some of the most precise measurements of solar and atmospheric neutrinos to date~\cite{Super-Kamiokande:2005wtt, Super-Kamiokande:2008ecj, Super-Kamiokande:2010tar}. Its successor, Hyper-K, is currently under construction and will reach a fiducial mass of 187~kton, significantly improving the sensitivity to MeV-GeV neutrino measurements~\cite{Abe:2011ts}.

Figure~\ref{fig:Obser} (right) shows the all-sky atmospheric muon neutrino background flux between 100~MeV and 100~GeV (solid purple), which originates primarily from cosmic-ray interactions in the atmosphere~\cite{Super-Kamiokande:2015qek}. To isolate the component that could be confused with solar neutrinos, I apply an energy-dependent angular cut based on the angular resolution of a water Cherenkov detector~\cite{Konishi:2010mv, Konishi:2011sc}. This resolution combines two energy-dependent contributions: an irreducible kinematic uncertainty arising from the angular distribution of muons produced in $\nu_\mu$ charged-current scattering, and a detector-level uncertainty associated with the reconstruction of muon tracks in Super-K~\cite{Galkin:2008qe, Super-Kamiokande:2007uxr}. The resulting ``fiducial'' atmospheric flux, shown as the dashed magenta curve, represents the residual background after restricting the dataset to events consistent with the direction of the Sun~\cite{Nguyen:2025ygc}.

The neutrino flux at Earth from millicharged DM annihilation into a specific SM final state is given by
\begin{equation}
    \frac{{\rm d}\Phi_{\nu_{\mu}}}{{\rm d}E_{\nu_{\mu}}}=\frac{\Gamma_{\chi^{+}\chi^{-}\to {\rm SM}}}{4\pi D_{\odot}^{2}}\,\frac{{\rm d}N_{\nu_{\mu}}}{{\rm d}E_{\nu_{\mu}}}\,P_{\rm surv}^{\nu}\,,
\end{equation}
where $D_{\odot}=1$~AU is the Earth-Sun distance and ${\rm d}N_{\nu_\mu}/{\rm d}E_{\nu_\mu}$ is the muon-neutrino spectrum per annihilation at Earth. For the $\tau^{+}\tau^{-}$ channel considered in this work, the neutrino yield is dominated by secondary production: the primary taus decay promptly via $\tau \to \nu_\tau + \text{hadrons/leptons}$, producing prompt $\nu_\tau$ along with secondary muons and charged pions, which subsequently decay through $\mu^\pm \to e^\pm \nu_e \nu_\mu$ and $\pi^\pm \to \mu^\pm \nu_\mu$ to generate additional neutrinos. I compute the resulting spectrum using the \texttt{PPPC4DM$\nu$} package~\cite{Baratella:2013fya}, which incorporates the full secondary cascade from these intermediate muons and pions, accounts for their cooling and energy loss in the dense solar environment before they decay, and provides the oscillation-averaged muon-neutrino flux at Earth\footnote{An updated treatment of the spectra is also available in the \texttt{$\chi$aro$\nu$} package~\cite{Liu:2020ckq}.}.

While most neutrinos escape the Sun freely, a small fraction can be absorbed through scattering with solar nuclei, predominantly hydrogen. The neutrino survival probability is~\cite{Berlin:2024lwe}
\begin{equation}
    P_{\rm surv}^{\nu}(E_{\nu})=\exp\left[-\int_{0}^{R_{\odot}}\!{\rm d}r\, \sigma_{\nu H}(E_{\nu})\,n_{H}(r) \right]\,,
\end{equation}
where $n_{H}(r)$ is the solar hydrogen density profile from the Standard Solar Model~\cite{Magg:2022rxb} and $\sigma_{\nu H}$ is the total neutrino-hydrogen cross section from Ref.~\cite{Zhou:2023mou}. The survival probability exceeds 99\% for $E_{\nu}\lesssim 4$~GeV and remains $\sim 96\%$ at $E_{\nu}=100$~GeV, so attenuation has only a modest effect on the signal.

\begin{figure*}[t!]
\centering
\includegraphics[width=2\columnwidth]{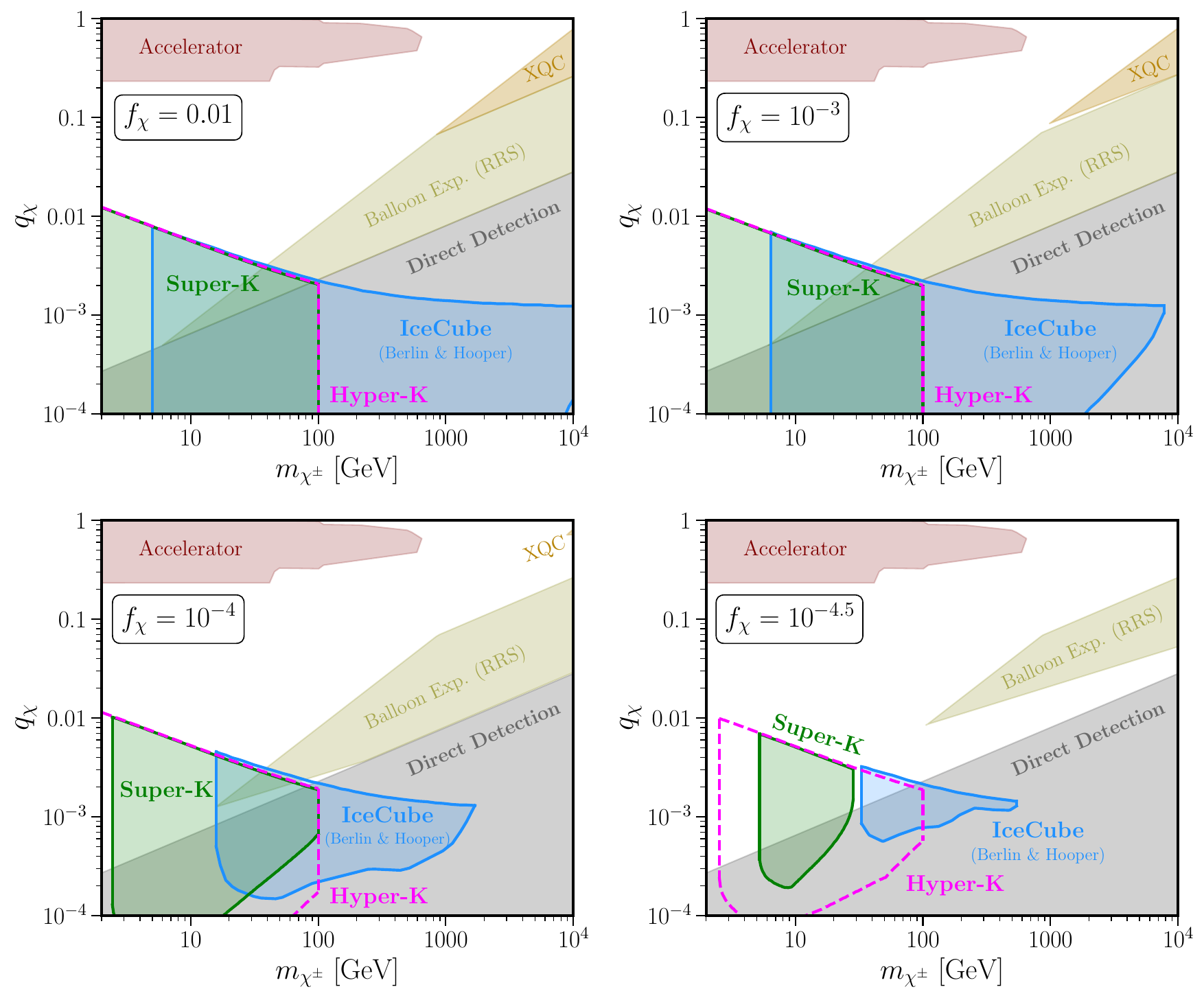}
\vspace{-0.5cm}
\caption{Solar neutrino constraints for millicharged DM annihilation to $\tau^{+}\tau^{-}$, with different DM fractions: IceCube results from Ref.~\cite{Berlin:2024lwe} are in blue, while Super-K constraints using 10-year observations are in green. The 10-year Hyper-K projections are in dashed magenta lines. Previous constraints include: direct detection (gray)~\cite{Emken:2019tni}, accelerator searches (maroon)~\cite{CMS:2012xi, Davidson:2000hf, CMS:2013czn, CMS:2024eyx}, rocket-based XQC experiment~\cite{Erickcek:2007jv, Mahdawi:2018euy} (gold), and the balloon-based RRS mission~\cite{Rich:1987st, Emken:2019tni} (olive)}
\label{fig:Compare}
\end{figure*}

Figure~\ref{fig:Obser} (right) shows the muon neutrino fluxes from millicharged DM annihilation into $\tau^{+}\tau^{-}$ inside the Sun, for a benchmark $m_{\chi}=10$~GeV and $q_{\chi}=10^{-3}$ (deep in the equilibrium regime), as a function of the DM fraction $f_{\chi}$ scanned from $10^{-3}$ down to $10^{-7}$. Even for fractions as small as $f_{\chi}\sim 10^{-5}$, the resulting solar neutrino flux from millicharged DM annihilation can exceed the fiducial atmospheric muon neutrino background, demonstrating that water Cherenkov detectors are sensitive to highly subdominant millicharged DM populations.

Using both the muon neutrino flux from DM annihilation and the fiducial atmospheric background, I compute the expected number of observed neutrino events in the energy range $[E_{\nu}^{\rm min}, E_{\nu}^{\rm max}]$ as

\begin{equation}
    N_{\nu}^{\chi/{\rm bkg}}=\int_{E_{\nu}^{\rm min}}^{E_{\nu}^{\rm max}}\!{\rm d}E_{\nu}\,\frac{{\rm d}\Phi_{\nu}^{\chi/{\rm bkg}}}{{\rm d}E_{\nu}}\,\sigma_{\nu {\rm H}_{2}{\rm O}}(E_{\nu})\times \xi\,,
\end{equation}

\noindent where $\sigma_{\nu {\rm H}_2{\rm O}}$ is the neutrino-water cross section from Ref.~\cite{Zhou:2023mou} and $\xi=N_{H_{2}O}\times T_{\rm obs}$ is the detector exposure, with $N_{H_{2}O}$ the number of water molecules in the fiducial volume and $T_{\rm obs}$ the observation time. Throughout this work, I assume an exposure of $T_{\rm obs}=10$~years for both detectors, with fiducial masses of 22.5~kton for Super-K and 187~kton for Hyper-K. For the $\tau^+\tau^-$ channel considered here, I integrate the spectrum from $E_{\nu}^{\rm min}=0.33\,m_{\chi}$ to $E_{\nu}^{\rm max}=1.2\,m_{\chi}$, which encompasses the bulk of the DM signal while limiting contamination from the atmospheric background~\cite{Nguyen:2025ygc}.

To set limits on the millicharged DM parameter space, I assume Poisson-distributed events and impose a 95\% confidence level (CL) exclusion criterion. A given parameter point is excluded if the predicted total event count $N_{\rm DM}+N_{\rm bg}$ is large enough that a background-only fluctuation producing this many events or more would occur with probability below 5\%, i.e.,
\begin{equation}
    \sum_{k=N_{\rm DM}+N_{\rm bg}}^{\infty}\frac{\lambda^{k}e^{-\lambda}}{\Gamma(k+1)}\leq 0.05\,,
\end{equation}
where $\lambda \equiv N_{\rm bg}(m_\chi)$ is the expected number of fiducial atmospheric muon-neutrino background events within the energy window $[E_\nu^{\rm min}, E_\nu^{\rm max}]=[0.33\,m_\chi, 1.2\,m_\chi]$ used for the $\tau^+\tau^-$ analysis. To avoid spurious limits in regions of parameter space where both the predicted signal and background are very small, I additionally require $N_{\rm DM}+N_{\rm bg}\geq 2.7$ for an exclusion to be claimed~\cite{Feldman:1997qc}. Solving these conditions jointly for each $(m_\chi, q_\chi, f_\chi)$ point yields the 95\% CL exclusion regions presented in the next section.

\section{Constraints on the millicharged particle}
\label{sect:constraint}

Figure~\ref{fig:Compare} shows the Super-K constraints (green shaded regions) and Hyper-K projections (magenta dashed lines), both assuming 10 years of solar muon-neutrino observation, for four representative values of the millicharged DM fraction $f_\chi$ across the $m_\chi=2\text{--}100$~GeV mass range. For comparison, I also display existing constraints in this strongly-coupled parameter space. The direct-detection ceiling is shown in gray~\cite{Emken:2019tni}, and combined constraints from collider experiments, primarily from the CMS collaboration~\cite{CMS:2012xi, Davidson:2000hf, CMS:2013czn, CMS:2024eyx}, are shown in maroon; both are independent of $f_\chi$. Additional constraints in the strongly-coupled regime, including the balloon-borne RRS experiment (olive)~\cite{Rich:1987st} and the rocket-based XQC mission (gold)~\cite{Erickcek:2007jv, Mahdawi:2018euy}, weaken rapidly with decreasing DM fraction. Most importantly, I overlay the IceCube constraints (blue) from Ref.~\cite{Berlin:2024lwe} (Berlin and Hooper, 2024), which probe the same millicharged-DM solar-capture scenario considered here, allowing for a direct comparison.

For a DM fraction of 1\% (Figure~\ref{fig:Compare}, upper left), where the XQC and RRS constraints remain relevant, the Super-K upper boundary, set by bound-state suppression with thorium, is similar to that of IceCube, excluding millicharges $q_{\chi}\lesssim 10^{-2}$ for $m_{\chi}\lesssim 100$~GeV. More importantly, Super-K extends the constraint down to $m_{\chi}=2$~GeV, while the IceCube analysis stops at 5~GeV. In addition to the evaporation cutoff discussed in Sec.~\ref{ssect:capt_nX_eva}, Ref.~\cite{Berlin:2024lwe} stops at 5~GeV because their analysis relies on IceCube collaboration limits for the $\tau^{+}\tau^{-}$ channel~\cite{IceCube:2016dgk, IceCube:2021xzo}, which themselves do not extend below 5~GeV~\cite{IceCube:2021xzo}. Super-K, with its sensitivity to solar neutrinos down to $\sim 100$~MeV, therefore naturally extends the accessible parameter space. Furthermore, for DM fractions down to $f_{\chi}=10^{-3}$ (upper right panel), the Super-K limit remains essentially unchanged near $m_{\chi}=2$~GeV, while the IceCube exclusion region shrinks toward higher masses, with its lower edge retreating to $m_{\chi}\simeq 6.5$~GeV. This robustness reflects the fact that the lower energy threshold of Super-K captures a larger portion of the neutrino spectrum produced by light millicharged DM, compensating for the reduced overall flux at small $f_{\chi}$.

\begin{figure*}[t!]
\centering
\includegraphics[width=1\columnwidth]{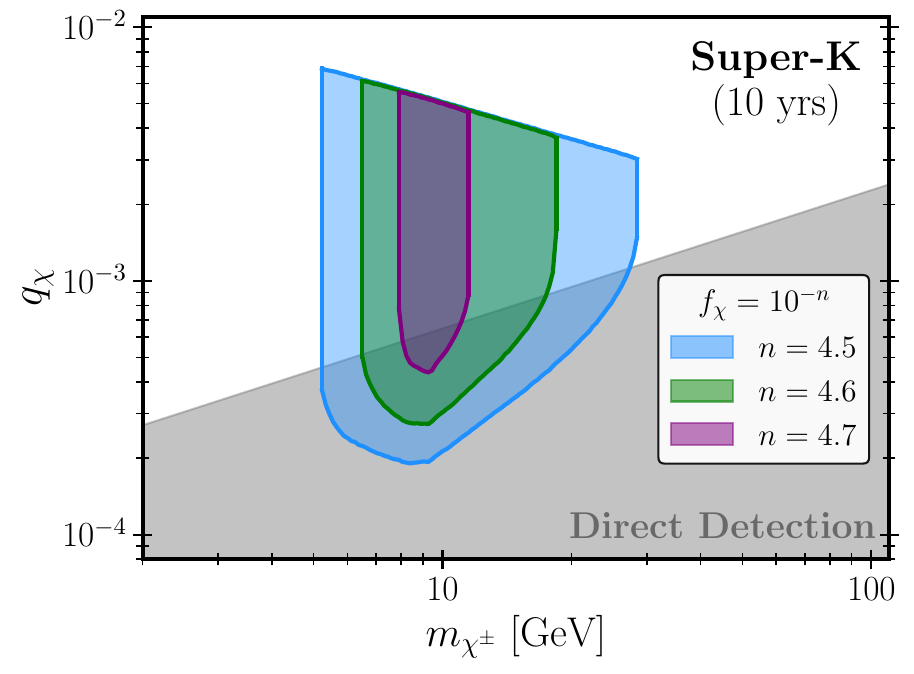}
\hfill
\includegraphics[width=1\columnwidth]{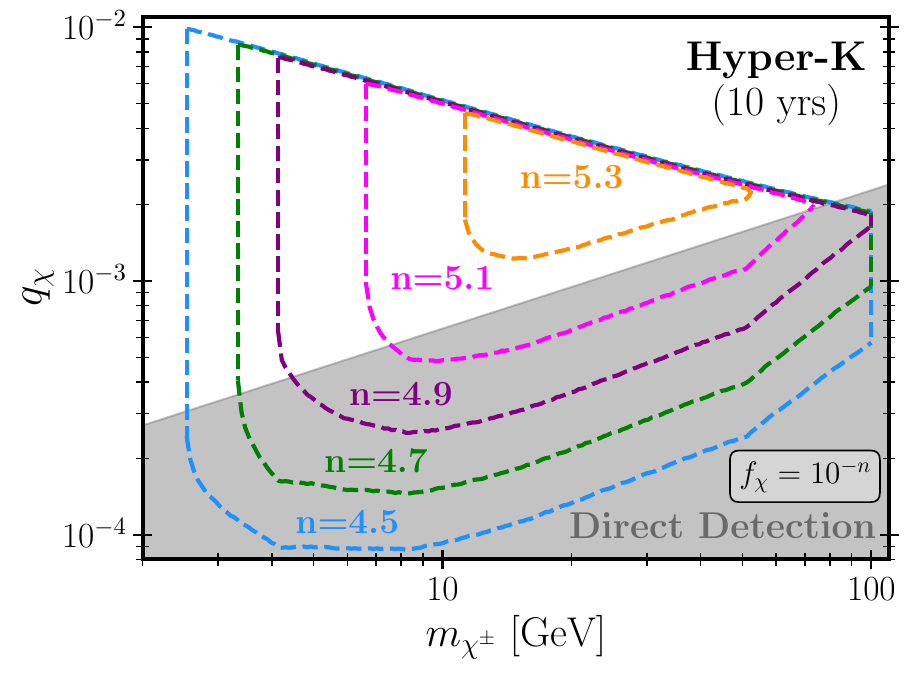}
\vspace{-0.5cm}
\caption{Water Cherenkov detector sensitivities for strongly-coupled millicharged DM for the small fraction of DM abundance, using 10-year neutrino observations from the Sun. {\bf Left:} Super-K constraints, that can be sensitive down to $f_{\chi}\simeq2\times10^{-5}$. {\bf Right:} Hyper-K projections, which can go down to $f_{\chi}\simeq 5\times 10^{-6}$. Upper limits from direct detection are in gray~\cite{Emken:2019tni}.}
\label{fig:SKHK}
\end{figure*}

At an even smaller DM fraction, $f_{\chi}=10^{-4}$, the difference between the Super-K and IceCube constraints begins to appear clearly. Super-K still covers the \mbox{$m_{\chi}=2.5\text{--}18$~GeV} window, which is inaccessible to IceCube. At higher masses, $m_{\chi}\gtrsim 30$~GeV, however, the Super-K lower bound on $q_{\chi}$ becomes weaker than the IceCube one, reflecting IceCube's superior sensitivity to high-mass DM. The projected Hyper-K reach (dashed magenta) closes this gap, surpassing IceCube across the entire $m_{\chi}\lesssim 100$~GeV range.

Notably, at the smallest DM fraction considered in Ref.~\cite{Berlin:2024lwe} with IceCube, Figure~\ref{fig:Compare} (bottom right) shows that Super-K can probe a complementary region of parameter space, covering \mbox{$m_{\chi}=5\text{--}28$~GeV} and \mbox{$q_{\chi}=(4\text{--}7)\times 10^{-3}$}. The Hyper-K projection further extends this reach down to $m_{\chi}=2.5$~GeV and bridges the small gap between the IceCube and Super-K excluded regions. Together, these results demonstrate that Cherenkov neutrino detectors using either ice or water as the medium form a comprehensive and powerful probe of millicharged DM annihilation in the Sun in the strongly-coupled regime.

Finally, Figure~\ref{fig:SKHK} demonstrates the smallest DM fractions of millicharged particles that Super-K and Hyper-K can probe with a 10-year observation timescale\footnote{These constraints are conservative, since Super-K has in fact been operating for nearly 30 years. I thank Bei Zhou for pointing this out during our discussion.}. Compared to the direct-detection ceiling, which is the only competing constraint at the small fractions relevant for this parameter space, Super-K extends the reach down to $f_{\chi}\simeq 10^{-4.7}\approx 2\times 10^{-5}$, more than a factor of two smaller than the IceCube limit. Hyper-K is projected to push this sensitivity further, reaching $f_{\chi}\simeq 10^{-5.3}\approx 5\times 10^{-6}$, roughly an order of magnitude below the IceCube result from Ref.~\cite{Berlin:2024lwe}. At the smallest accessible fractions, the excluded regions narrow to a window where the combination of efficient capture and minimal bound-state suppression maximizes the signal. Below the lower edges of these regions, the predicted neutrino flux falls below detector sensitivity, while above the upper edges, bound-state formation with thorium nuclei kills the annihilation rate. The peak sensitivity of Super-K sits around $m_{\chi}\sim 10$~GeV, while Hyper-K's deepest reach shifts to $m_{\chi}\sim 20$~GeV, reflecting its improved sensitivity at higher neutrino energies thanks to the larger fiducial volume. Together, Hyper-K both extends the lower boundary downward by an order of magnitude in $f_{\chi}$ and broadens the accessible mass range.

\section{Conclusion and Outlook}
\label{sect:conclusion}

Millicharged DM in the strongly-coupled regime, where the cross section with baryonic matter exceeds the transition cross section of the Sun, can be efficiently captured and annihilate to produce a flux of high-energy neutrinos at Earth. Building on the recent IceCube analysis of Berlin and Hooper~\cite{Berlin:2024lwe}, in this work I have shown that water Cherenkov neutrino detectors, namely Super-Kamiokande and the upcoming Hyper-Kamiokande, provide a powerful and complementary probe of this scenario. The lower energy threshold of these detectors enables sensitivity to neutrinos from the annihilation of lighter millicharged DM, extending the accessible mass range down to $m_{\chi}=2$~GeV and filling the low-mass gap left by IceCube.

Using a 10-year solar neutrino observation timescale, I find that Super-K can constrain previously unexplored parameter space at $m_{\chi}\sim2\text{--}30$~GeV, complementing IceCube at higher masses. At a DM fraction of $f_{\chi}=10^{-2}$, the Super-K upper boundary set by bound-state suppression with thorium nuclei is similar to that of IceCube, excluding millicharges $q_{\chi}\lesssim 10^{-2}$ for $m_{\chi}\lesssim 100$~GeV, while the lower mass reach extends down to $m_{\chi}=2$~GeV, well below the IceCube cutoff at $5$~GeV. As $f_{\chi}$ decreases, the IceCube exclusion region progressively retreats toward higher masses, while the Super-K limit remains robust at low masses thanks to its lower energy threshold. At $f_{\chi}=10^{-4.5}$, Super-K is the only neutrino-telescope probe of millicharged DM in the $m_{\chi}\sim5\text{--}28$~GeV window.

The 10-year Hyper-K projection further extends this sensitivity. Hyper-K surpasses IceCube across the entire $m_{\chi}\lesssim 100$~GeV range and bridges the small gap that remains between the IceCube and Super-K excluded regions at small $f_{\chi}$. The peak of Hyper-K's sensitivity shifts to $m_{\chi}\sim 20$~GeV, compared to $m_{\chi}\sim 10$~GeV for Super-K, reflecting its enhanced reach at higher neutrino energies thanks to the larger fiducial volume. In terms of fractional abundance, Super-K probes down to $f_{\chi}\simeq 2\times 10^{-5}$, more than a factor of two smaller than IceCube, while Hyper-K reaches $f_{\chi}\simeq 5\times 10^{-6}$, roughly an order of magnitude below the IceCube limit. Together, these Cherenkov neutrino detectors, using either ice or water as the medium, form a comprehensive and powerful probe of millicharged DM annihilation in the Sun in the strongly-coupled regime, accessing parameter space that is invisible to direct detection, accelerator searches, and the strongly-interacting probes from XQC and RRS.

The results presented here motivate the search for solar millicharged DM at upcoming large neutrino detectors, including Hyper-K~\cite{Abe:2011ts}, KM3NeT~\cite{KM3NeT:2024xca}, TRIDENT~\cite{TRIDENT:2022hql}, IceCube-Upgrade~\cite{Ishihara:2019aao}, and IceCube-Gen2~\cite{IceCube-Gen2:2020qha}, both as improvements over current sensitivity and as complementary probes of DM masses below and above the electroweak scale. More broadly, they reinforce the case for neutrino astronomy as a powerful tool not only for studying neutrino physics and astrophysical sources, but also for probing the nature of DM. Because the sensitivity to DM signals improves with the integrated exposure, dedicated DM searches provide additional scientific motivation to extend the operational lifetimes of these neutrino observatories well beyond their original mission timescales.

Beyond direct improvements to the analysis presented here, several complementary observational strategies naturally follow from the assumptions and limitations of this study:
\begin{itemize}
    \item Other annihilation channels of the dark photon mediator, such as $\mu^{+}\mu^{-}$ and $\pi^{+}\pi^{-}$, become relevant for DM masses below the tau threshold. The neutrino spectra from these channels are shifted to lower energies by muon and pion cooling inside the Sun~\cite{Rott:2012qb, Bernal:2012qh}, making them accessible to MeV-scale solar neutrino observations at Super-K, Borexino, and JUNO. 
    \item In this work, as well as in Ref.~\cite{Berlin:2024lwe}, the dark coupling $\alpha_{D}$ is taken to be very small, so that the $\chi^{+}\chi^{-}\to A^{\prime}A^{\prime}$ channel is negligible. However, this channel can be probed in the regime where the long-lived dark photon escapes the Sun and decays through the dark photon trident process $A^{\prime}\to 3\gamma$~\cite{Linden:2024uph}, producing a signal accessible to solar $\gamma$-ray observations~\cite{Linden:2020lvz, Leane:2017vag, Linden:2025xom}.
    \item In addition to the Sun, Jupiter is an attractive target for lighter millicharged DM. Its evaporation mass is lower than that of the Sun: around $\sim 1$~GeV for nuclear scattering~\cite{Garani:2021feo, Robles:2024tdh}. Existing gamma-ray~\cite{Leane:2021tjj} and neutrino~\cite{Robles:2024tdh} observations of Jupiter could already provide meaningful constraints in this regime.
\end{itemize}
These directions, which target lighter millicharged DM, require dedicated studies of capture and evaporation with an ultralight dark photon mediator, especially below the solar evaporation mass. Together with the present analysis, they form a coherent program of indirect searches for strongly-coupled millicharged DM using neutrino and $\gamma$-ray observations of celestial objects. I leave these explorations for future work.
\vspace{0.3cm}
\section*{Acknowledgement}
\vspace{-0.3cm}
I thank Marianne Moore and Bei Zhou for fruitful discussions. I thank Marianne Moore, Isabelle John, Tim Tait and Tim Linden for comments on the draft. I am especially grateful to Dan Hooper for explaining his work~\cite{Berlin:2024lwe} during the IDM conference in L'Aquila, Italy, and to Asher Berlin for kindly answering my questions on millicharged DM and the IceCube analysis. I thank Carlos Blanco and the Institute for Gravitation and the Cosmos (IGC) at Penn State University for their hospitality, and my advisor, Tim Linden, for his encouragement in finishing this paper. This work is supported by the Swedish Research Council under contract 2022-04283, and by two grants from the Royal Swedish Academy of Sciences (KVA): PH2025-0073 (Physics) and AST2025-0048 (Astronomy and Space Science). I acknowledge the support from EDUCATE Excellence Centre funded by the Swedish Research
Council through grant Dnr~2022-06627.

\noindent {\bf AI Usage Statement.} Claude Code was used to assist with generating the plots. Claude was used for grammar and typos checks. AI tools were not used to generate scientific results or draw conclusions. All analysis, code, and the final text were produced, checked, and verified by the author, who takes the full responsibility for the content of this work.

\bibliography{ref}
\end{document}